\documentclass[]{raa_rb}

\usepackage{graphicx,times}             
\usepackage{natbib}

\usepackage{amsmath}
\usepackage{booktabs,longtable}
\usepackage{pdflscape}
\newcommand{\bd}{\begin{displaymath}}
\newcommand{\ed}{\end{displaymath}}

\newcommand{\mgii}{Mg\,{\sc ii}}
\newcommand{\civ}{C\,{\sc iv}}
\newcommand{\hb}{H{\sc $\beta$}}
\newcommand{\hg}{H{\sc $\gamma$}}

\begin{document}

\title{Constraints on black hole spins with a general relativistic accretion disk corona model}

   \volnopage{Vol.0 (200x) No.0, 000--000}      
   \setcounter{page}{1}          

   \author{Bei You\inst{1,3}, Xinwu Cao\inst{1}, and Ye-Fei Yuan\inst{2}
   }

   \institute{Key Laboratory for Research in Galaxies and
Cosmology, Shanghai Astronomical Observatory, Chinese Academy of
Sciences, 80 Nandan Road, Shanghai, 200030, China; {\it youbeiyb@gmail.com, cxw@shao.ac.cn}\\
        \and
             Department of Astronomy, University of Sciences and
Technology of China, Hefei, Anhui 230026, China; {\it yfyuan@ustc.edu.cn} \\
        \and
             University of Chinese Academy of Sciences, 19A
Yuquanlu, Beijing 100049, China \\  
}

\abstract{
The peaks of the spectra of the accretion disks surrounding massive
black holes in quasars are in the far-UV or soft X-ray band, which
are usually not observed. However, in the disk corona model, the
soft photons from the disk are Comptonized to high energy in the hot
corona, and the hard X-ray spectra (luminosity and spectral shape)
contain the information of the incident spectra from the disk. The
values of black hole spin parameter $a_{\ast}$ are inferred from the
spectral fitting, which spread over a large range, $\sim -0.94$ to
$0.998$. We find that the inclination angles and mass accretion
rates are well determined by the spectral fitting, while the results
are sensitive to the accuracy of black hole mass estimates.
No tight constraints on the black hole spins are achieved,
if the uncertainties of black hole mass measurements are a factor of
four, which are typical for the single-epoch reverberation mapping
method. Recently, the accuracy of black hole mass measurement has
been significantly improved to $0.2-0.4$~dex with velocity resolved
reverberation mapping method\citep*[][]{2014MNRAS.445.3073P}. The
black hole spin can be well constrained if the mass measurement
accuracy is $\la 50$\%. In the accretion disk corona scenario, a
fraction of power dissipated in the disk is transported into the
corona, and therefore the accretion disk is thinner than a bare disk
for the same mass accretion rate, because the radiation pressure in
the disk is reduced. We find that the thin disk approximation, 
$H/R\la 0.1$, is still valid if $0.3<\dot{m}<0.5$, provided a half of
the dissipated power is radiated in the corona above the disk.
\keywords{Quasars: accretion disk; X-ray: corona; Black hole physics; Galaxies: active}
}
   \authorrunning{B. You et al. }            
   \titlerunning{Constraints on black hole spins with a general relativistic accretion disk corona model}  
\maketitle

\section{Introduction}

It is believed that black holes reside in X-ray binaries and active
galactic nuclei (AGN) with typical masses $\sim10M_{\odot}$ and
$10^{6}-10^{9}M_{\odot}$ respectively
\citep{1995ARA&A..33..581K,2000ApJ...539L...9F,2006ARA&A..44...49R,2008ApJS..175..356H,2012NewAR..56...93A}.
Massive black holes may grow up through gas accretion or/and
mergers, which is also related with the galaxy evolution
\citep*[e.g.,][]{1996MNRAS.283..854M,2004MNRAS.351..169M,2005ApJ...620...69V,2008ApJ...684..822B,2013ApJ...762...68D}.
The structure and radiation of accretion disks surrounding black
holes have been extensively studied in the last several decades
\citep{1973A&A....24..337S,1992apa..book.....F,2004MNRAS.353.1035M}.
Relativistic jets have been observed in some X-ray binaries and
AGNs, which is thought to be related to accretion disks or/and
spinning black holes
\citep{1977MNRAS.179..433B,1982MNRAS.199..883B}.

A black hole can be described by two quantities: the mass $M$
and the dimensionless spin parameter $a_{\ast}=cJ/GM^{2}$,
where $J$ is the angular momentum of the black hole. In the last
decade, great progresses have been achieved on the mass measurement
of black holes in either active or nonactive galaxies. The black
hole mass in nearby galaxies can be estimated with the correlations
between black hole mass and the stellar velocity dispersion or bulge
luminosity of the host galaxy
\citep{1995ARA&A..33..581K,2000ApJ...539L...9F,2011Natur.480..215M}.
For active galaxies, the size of the broad-line region (BLR) is
measured with reverberation mapping method, and the black hole mass
can be estimated with the broad-line width assuming the motion of
BLR clouds to be virialized \citep{2000ApJ...533..631K}. There is a
tight correlation between the BLR size and the optical continuum
luminosity, which is widely used to estimate masses of black holes
in AGNs, or called as single epoch reverberation mapping method
\citep*[e.g.,][]{2000ApJ...533..631K,2002ApJ...581L...5W,2011ApJS..194...45S,2012ApJS..201...38T}.

Compared with the mass estimates, the measurements of the black hole
spins are still far from satisfactory, though great efforts have
been devoted on this issue. The observed broad Fe K$\alpha$ emission
lines provide a way to estimate the black hole spin parameters,
which is applicable for black holes in both galactic X-ray binaries
and AGNs. The observed broad Fe K$\alpha$ emission lines are
believed to be emitted from the inner regions of the accretion
disks. The radii of the innermost stable circular orbits (ISCO) of
the gas in the accretion disks can be inferred from the observed
asymmetric Fe k$\alpha$ line profiles
\citep{2002A&A...387..215M,2002ApJ...577L..15M,2004MNRAS.351..466M,2006ApJ...652.1028B,2012ApJ...758...67L},
and the dimensionless spin $a_{\ast}$ can be determined with the
radii of the ISCO
\citep{1985MNRAS.216P..65B,2003PhR...377..389R,2006ARA&A..44...49R}.
The application of this method requires high quality X-ray
spectroscopic data, and a few black hole spins in nearby
Seyfert galaxies have been measured in this way
\citep{2006A&A...454..741G,2006ApJ...652.1028B,2010A&A...524A..50D,2011MNRAS.411.2353P,2012ApJ...758...67L}.

The black hole spin parameters in some X-ray binaries were measured
by fitting the observed continuum spectra with general relativistic
accretion disk models \citep*[so-called ``continuum-fitting
method",][]{1997ApJ...482L.155Z,2006ApJ...636L.113S,2009ApJ...701L..83S,
2011ApJ...742...85G,2011CQGra..28k4009M,arXiv:1506.03959}.
In principle, this method can be used to constrain the black hole
spins for AGNs. In some early works, the accretion disk models were
used to fit the observed multi-band spectra of quasars
\citep*[e.g.,][]{1987ApJ...314..699B,1989ApJ...346...68S}, in which
the general relativistic effects have been taken into account
\citep{1975ApJ...202..788C,1991ApJ...376...90L}.
{\cite{2011MNRAS.415.2942C} estimated the black hole spin to be
$a_{\ast}=0.3$ by fitting the broad band (IR/optical/UV) continuum spectrum of
the quasar SDSS J094533.99+100950.1.} It was found that the model
fitting suffers from the degeneracy of the disk parameters, namely,
the black hole mass $M$, the spin $a_{\ast}$, and the mass
accretion rate $\dot{M}$. At the time before the invention of the
black hole mass measurements used nowadays, the derived disk
parameters with the spectral fittings are quite uncertain. It is
possible to constrain the the spins of the massive black holes in
AGNs, when the black hole masses are measured. The degeneracy of the
parameters in accretion disk spectral fittings can be eliminated if
the peak frequency and luminosity of the accretion disk are well
observed. However, the application of this method in AGNs meets a
major difficulty, i.e., the spectral peaks of accretion disks
surrounding massive black holes with $\sim10^{8}M_{\odot}$ are in
the far-UV or soft X-ray band, which are unobservable for most AGNs.

The observed ultraviolet (UV)/optical emission of AGNs is thought to
be a thermal emission from the optically thick accretion flows
\citep{1978Natur.272..706S,1982ApJ...254...22M,1989ApJ...346...68S},
while they are too cold to reproduce the observed power-law hard
X-ray spectra of AGNs. The hard X-ray spectra of AGNs
most likely originate from the inverse Compton process in which soft photons
from the disks are scattered by the hot electrons in the coronas
\citep{1979ApJ...229..318G,1991ApJ...380L..51H,1998MNRAS.299L..15D}. According to this disk-corona
scenario, most gravitational energy is generated in the cold disk,
while a fraction of it is carried into the corona likely
by the magnetic loops in the disk
\citep{1998MNRAS.299L..15D,2002ApJ...572L.173L,2009MNRAS.394..207C}.
\citet{2009MNRAS.394..207C} calculated the structure and spectrum of
a geometrically thin accretion disk with a corona in AGNs,
in which the corona is assumed to be powered by the
reconnection of the magnetic loops originating from
the cold accretion disk due to buoyancy
instability. It is found that both the hard X-ray
bolometric correction factor $L_{\rm bol}/L_{\rm X,2-10 keV}$
and the hard X-ray spectral index
increase with the Eddington ratio, which are qualitatively
consistent with the observations
(e.g.,
Wang et al., 2004; Shemmer et al., 2006; Vasudevan \& Fabian, 2007; 
Zhou \& Zhao 2010; Fanali et al., 2013).
This model was expanded in the general relativistic frame for a spinning Kerr black hole by
\citet{2012ApJ...761..109Y}, in which the spectra of such a
accretion disk-corona are recalculated taking into account general relativistic effect with the
relativistic ray-tracing method.

As discussed above, the peaks of the spectra of the accretion disks
surrounding massive black holes in AGNs are usually not observed
\citep*[but also see][for the case of a relative small black hole in
a narrow-line quasar]{2010ApJ...723..508Y}. However, a portion of
the soft photons originating from the disk undergoing Comptonization in
the hot corona, and the scattered photons are dominantly in the hard
X-ray band. This implies that the observed hard X-ray spectra
(luminosity and spectral shape) contain the information of the
incident spectra from the cold disk. It seems possible to constrain
the values of black hole spin parameter $a_{\ast}$ in AGNs with a general
relativistic accretion disk corona model, as a variant of the continuum-fitting method, even if the spectral peaks
of the accretion disks are not observed.

In this paper, we use the general relativistic model of an accretion
disk-corona surrounding a spinning black hole to fit the
observed multi-band spectra of AGNs with measured black hole masses.
The black hole spinning parameter $a_{\ast}$ are tentatively
derived for a small sample of AGNs. We briefly summarize the
accretion disk-corona model used for AGN spectral fitting in Section
2. The results and discussion are in Sections 3 and 4. Throughout
this paper, a cosmology with $H_{0}=70 {\rm km~ s}^{-1}{\rm
Mpc}^{-1}$, $\Omega_{\rm m}=0.3$ and $\Omega_{\Lambda}=0.7$ is
adopted.

\section{Model}

In this work, we adopt the Kerr black hole accretion disk-corona model
developed by
\citet{2012ApJ...761..109Y}. All the general relativistic effects
are properly considered in both the calculations of the structure
and emergent spectra of the accretion disks. We briefly summarize
the model in the next two sub-sections \citep*[see][for the
details]{2012ApJ...761..109Y}.

\subsection{Structure of an accretion disk with corona}

In the accretion disk corona model, most gravitational energy of the
gas is dissipated in the cold disk, and a portion of it is
carried into the corona by the magnetic loops. The particles in
the corona are heated by the re-connection of the loops in the
corona \citep{1998MNRAS.299L..15D}. The electrons in the corona are
hot, and the observed hard X-ray spectra mostly originate from the
inverse Compton process in which soft photons from the thermal radiation
of the cold disk are scattered to high energy by the
hot electrons
\citep*[e.g.,][]{1979ApJ...229..318G,1991ApJ...380L..51H,1994ApJ...436..599S}.

The gravitational power of the gas generated in the cold accretion disk surrounding a spining black
hole is
\begin{equation}\label{q_disk_plus}
Q_{\rm dissi}^{+}={\frac {3GM\dot{M}}{8\pi R^{3}}}{\frac {L}{BC^{1/2}}},
\end{equation}
where $L$, $B$, $C$ are the
general relativistic correction factors
\citep{1973blho.conf..343N,1974ApJ...191..499P}, and
$\dot{M}$ is the mass accretion rate of the disk.

The fraction of the dissipated power transported to the corona can
be estimated provided the strength of the magnetic field is known.
The magnetic pressure of the disk is usually assumed to scale with
the gas or/and radiation pressure of the disk depending on the
different magnetic stresses adopted
\citep{1981ApJ...247...19S,1984ApJ...277..312S,1984ApJ...287..761T,2009MNRAS.394..207C,2009ApJ...704..781H}.
The detailed physics of the field generation is still quite unclear,
though the different magnetic stressed are tested against the
observations in AGNs in the previous works
\citep{2004ApJ...607L.107W,2009MNRAS.394..207C,2012ApJ...761..109Y}.
For simplicity, the ratio $f$ of the dissipated power in the corona
to that in the disk is taken as a parameter in our model
calculations,
\begin{equation}\label{q_cor_plus}
  f={\frac {Q_{\rm cor}^{+}}{Q_{\rm dissi}^{+}}},
\end{equation}
which will be determined by the comparison with the observed
spectra.

A portion of the soft photons from the disk undergo inverse Comptonization
to high energy in the corona, which correspondingly cools
the electrons in the corona. About half of the
scattered photons are intercepted by the disk, some of which are reflected
and the rest heats the disk and re-radiate as blackbody radiation
\citep{1999MNRAS.303L..11Z}. Therefore the energy equation for the cold
disk is
\begin{equation}\label{energy_equation}
Q_{\rm dissi}^{+}-Q_{\rm cor}^{+}+\frac{1}{2}(1-\varepsilon)Q_{\rm
cor}^{+}=\frac{4\sigma T_{\rm d}^{4}}{3\tau}=\sigma T_{\rm s}^{4},
\end{equation}
where the
reflection albedo $\varepsilon=0.15$ is adopted in all our
calculations \citep{1999MNRAS.303L..11Z}, $\tau=\tau_{\rm es}+\tau_{\rm ff}$ is the
optical depth in the vertical direction of the disk, and $T_{\rm d}$ and $T_{\rm s}$
are the temperature in the mid-plane of the disk and the temperature of the gas at the disk
surface respectively.

\subsection{Emergent spectrum of the accretion disk with corona}

It is well known that most of the radiation of the disk comes from
the inner region of the disk \citep{1973A&A....24..337S}. The
spectrum of an accretion disk-corona for a spinning
black hole is influenced by the relativistic effects, i.e., the
frame dragging, gravitational redshift, and photon trajectory
bending, due to the strong gravity field of the black hole
\citep*[][]{1985Ap&SS.113..181Z,1989MNRAS.238..897L,1991ApJ...376...90L,2005ApJS..157..335L,2009ApJ...699..513L,2012ApJ...761..109Y},
especially for the emission from the region near the black hole. The
ray-tracing method is widely adopted in this kind of spectral
calculations
\citep{1989MNRAS.238..897L,1991ApJ...376...90L,1997ApJ...482L.155Z,2005ApJS..157..335L,2012ApJ...761..109Y}.
Considering the complexity of the radiation transfer of Comptonized
photons in the geometrically thick corona,
\cite{2012ApJ...761..109Y} calculated the emergent Comptonized
spectrum from the corona by dividing layers in the corona with the
ray-tracing method. In this work, we follow the same routine of
\cite{2012ApJ...761..109Y} to calculate the spectrum of an
accretion disk-corona for a rotating black hole, in
which all the general relativistic effects are taken into account
\citep*[see Section 3.4 in][for the details]{2012ApJ...761..109Y}

The radiation from geometrically thin, optically thick accretion
disk is anisotropic, and the specific intensity of the disk
emission, $I_{\mu}\propto (1+2\mu)$, in the co-moving frame of the
disk, where $\mu=\cos\theta$ \citep*[$\theta$ is the angle between
the line of sight and the normal of the
disk][]{1960ratr.book.....C,1985A&A...143..374S,2011ApJ...728L..44L}.

UV/soft X-ray photons are emitted from the inner region of
the accretion disk, where a hot plasma layer is present above the
disk surface caused by the illumination of the external hard X-ray
emission from the corona. The electron scattering opacity dominates
over absorption in this hot layer, and the emergent spectrum of the
disk departures from the black body emission due to radiative
transfer. \cite{2001ApJ...559..680H} calculated the vertical
structure and emergent spectra of the disks with a hot plasma layer
by including a self-consistent treatment of Compton scattering.
\cite{2002ApJ...572...79C} found that the derived emergent spectrum
could be well reproduced by blackbody emission with a higher color
temperature, $T_{\rm rad}=f_{\rm col}T_{\rm d}$. The empirical color
correction factor $f_{\rm col}$ is given by
\begin{equation}\label{color_correction}
    f_{\rm col}(T_{\rm d})=f_{\infty}-\frac{(f_{\infty}-1)[1+\exp(-\nu_{\rm b}/\Delta\nu)]}{1+\exp[(\nu_{\rm p}-\nu_{\rm b})/\Delta\nu]}
\end{equation}
where $\nu_{\rm p}\equiv2.82k_{\rm B}T_{\rm d}/h$,
$f_{\infty}=2.3$ and $\nu_{\rm b}=\Delta\nu=5\times10^{15} \rm Hz$.
The local spectrum emitted from the disk can be calculated with
\begin{equation}\label{UV}
    F_{\nu}\propto \frac{1}{f_{\rm col}^{4}} \pi I_{\nu}(f_{\rm col}T_{\rm
    d}),
\end{equation}
where $I_{\nu}$ is the local blackbody emission.

\begin{figure}
   \centering
   \includegraphics[width=11.0cm, angle=0]{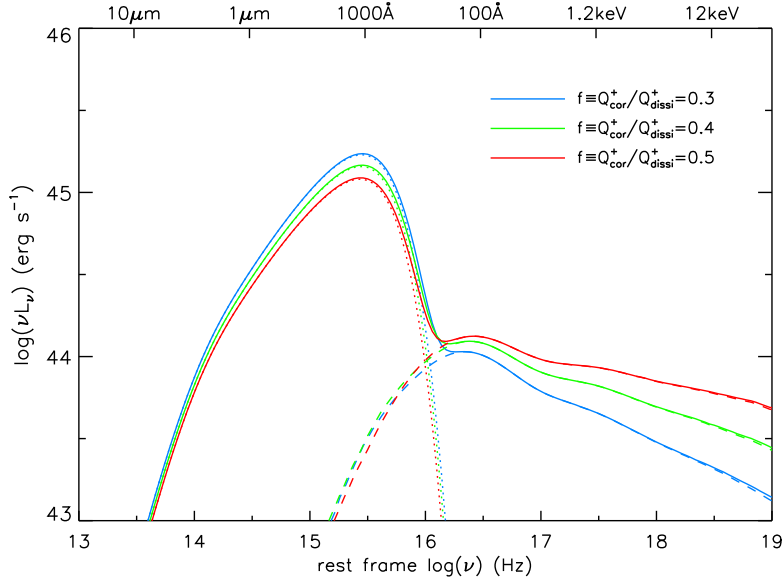}
\caption{The representtive spectra of the disk-corona model with different values of the ratio $f$ of the dissipated power in the corona to that in the disk. The dotted lines represent the spectra of the disk blackbody radiation, while the dashed lines represent the Compton radiation from the corona. The synchrotron and bremsstrahlung radiation which dominate
in radio bands are not plotted. It is found that the hard X-ray spectra become harder with the ratio $f$.}
\label{f_sed}
\end{figure}

As described in Section 2.1, the structure of the accretion disk-corona
can be calculated when the
values of the black hole mass $M$, the dimensionless spin
$a_{\ast}$, the mass accretion rate $\dot{m}$, and the ratio $f$
of the dissipated power in the corona to that in the disk, are
specified. With the derived structure of the accretion disk, we use
the ray tracing method to calculate the emergent spectrum of the
disk corona system observed in infinity at a viewing angle $\theta$. We plot the representive emergent spectra with different values of the ratio $f$ in Fig.~\ref{f_sed} to show the effect of $f$ on the overall shape of the spectra.  At a given radius, both the energy dissipated in the disk and the disk temperature decrease with energy ratio $f$ according to Eq.~(\ref{q_cor_plus}). Therefore, the specific intensity of the soft photons from the disk will also decrease with the ratio $f$. In this case, the electron temperature will increase with $f$, since the cooling of the corona is dominated by the inverse Compton scattering of the soft photons from the disk by the hot electrons in the corona \citep{2009MNRAS.394..207C,2012ApJ...761..109Y}. This means that the resulting spectra in the hard X-ray band become much harder and brighter with the increase in the ratio $f$, given that more soft photons from the disk will be Compton scattered to higher energy by the hot electrons in the corona (see Fig.~\ref{f_sed}). 

\section{Results}

Using the accretion disk model described in Section 2, we fit the
observed multi-band spectral energy distributions (SEDs) of five quasars which exhibit obvious
feature of so-called ``Big Blue Bump" (BBB) with measured black hole masses and good X-ray
measurements.
The peak of BBB at
$\sim 10^{15}-10^{16}$~Hz and a power-law hard X-ray spectrum are
the typical features of the spectrum of an accretion disk with
corona \citep*[e.g.,][]{2009MNRAS.394..207C,2012ApJ...761..109Y}. As
discussed in Section 1, the peaks of the BBBs are beyond the
waveband coverage of the instruments for most AGNs due to their
massive black holes. However, the left tail of the BBBs can be
observed in infared/optical/UV wavebands, which provide useful
information of the BBBs \citep*[e.g.,][]{2013MNRAS.431..210C}.
The SED data of five quasars in this work is taken from \citet{2011ApJS..196....2S} and summarized in Tables 1 and 2. 

\begin{table}
\begin{center}
\caption{The SED data in
optical/UV wavebands\label{sample}}
\setlength{\tabcolsep}{2.5pt}
\begin{tabular}{lccccccc}
\hline\noalign{\smallskip}
Object & Other name& R.A.(J2000)& Decl.(J2000)& z & $E(B-V)^{\rm a}$ & 
$\lambda L_{\lambda}(3000{\AA})^{\rm b}$& SampleID  \\
\hline\noalign{\smallskip}
PG 1322+659&          &13:23:49.54 & $+$65:41:48.0 & 0.1684 & 0.019 & 44.61 & RQ\\
PG 1115+407&          &11:18:30.20 & $+$40:25:53.0 & 0.1541 & 0.016 & 46.53 & RQ\\
4C 10.06& PKS 0214+10 &02:17:07.66 & $+$11:04:10.1 & 0.4075 & 0.109 & 45.54 & RL\\
4C 39.25& B2 0923+39  &09:27:03.01 & $+$39:02:20.9 & 0.6946 & 0.014 & 45.70 & RL\\
OS 562&               &16:38:13.45 & $+$57:20:24.0 & 0.7506 & 0.013 & 43.51 & RL\\
\hline\noalign{\smallskip}
\end{tabular}
\end{center}
\tablecomments{0.86\textwidth}{$^{\dag}$ (a) From
NED(http://nedwww.ipac.caltech.edu/) based on
\cite{1998ApJ...500..525S}.  (b) the rest-frame luminosity at
3000${\rm \AA}$. }\label{sample}
\end{table}

\begin{table}
\begin{center}
\caption[]{The X-Ray Spectral data}
\setlength{\tabcolsep}{2.5pt}
\begin{tabular}{lcccccccc}
\hline\noalign{\smallskip}
Object& E1& E2&
E3& \multicolumn{4}{c}{$f_{\nu}=f_{0}E^{\alpha}$}&
Reference\\
\cline{5-8}
&
&
&
&
$f_{0}$&
$\alpha$&
$f_{0}$&
$\alpha$&
\\
(1)&(2)&(3)&(4)&(5)&(6)&(7)&(8)&(9)\\
\hline\noalign{\smallskip}
PG 1322+659& 0.3 & 1.62& 10.0& 6.51E-4 & $-2.01^{+0.24}_{-0.11}$ &4.36E-4 &$-1.18^{+0.14}_{-0.11}$ & X,Po04\\
PG 1115+407& 0.3 & 2.04& 10.0& 6.48E-4 & $-1.85^{+0.06}_{-0.02}$ &4.05E-4 &$-1.19^{+0.10}_{-0.10}$ & X,Po04\\
4C 10.06& 0.1 & & 2.4& 9.08E-4 & $-1.13^{+0.52}_{-0.56}$ & & & R,Br97\\
4C 39.25& 0.1 & & 2.4& 5.83E-4 & $-1.25^{+0.06}_{-0.06}$ & & & R,Br97\\
OS 562& 0.1 & & 2.4& 1.36E-4 & $-1.38^{+0.04}_{-0.04}$ & & & R,Br97\\
\hline\noalign{\smallskip}
\end{tabular}
\end{center}
\tablecomments{0.86\textwidth}{$^{\dag}$For the sources PG~1322+659 and PG~1115+407, their
X-ray spectra are fitted by a broken power law model in the ranges
of $E1-E2$ and $E2-E3$ (keV) in the observer's frame. The remainders
are fitted by a single power law model between $E1$ and $E3$. The
columns (5) and (7) are the flux densities at 1~keV in units of mJy
($10^{-26} {\rm erg~ s^{-1}cm^{-2} Hz^{-1}}$); $E$ in keV.}
\tablerefs{0.86\textwidth}{ R,C,X indicate data sources, corresponding to ROSAT,
Chandra, and XMM, respectively. Br97: \cite{1997A&A...319..413B};
Po04: \cite{2004A&A...422...85P} }
\end{table}



The black hole masses $M$ of these five quasars are given in
\citet{2012ApJS..201...38T}, which are estimated with the broad-line
width and the scaling relationships
\citep{2006ApJ...641..689V,2009ApJ...699..800V}. The infrared
emission from these quasars may probably be dominated by the
radiation from the dust tori, while the origin of the soft X-ray
excess observed in AGNs is still controversial \citep*[see, e.g.,
][and the references therein]{2012MNRAS.420.1848D}. Therefore, the
infrared and soft X-ray spectral data are not used in our model
fitting on the SEDs.

\begin{table}
\begin{center}
\caption{The fitting parameters of the sources \label{fitting}}
\setlength{\tabcolsep}{0.1in} 
\begin{tabular}{cccccccc}
\hline\noalign{\smallskip}
Object& log$M$&
$f$&$a_{\ast}$& $\dot{m}$& $\theta$ & $T_{e}(\rm keV)$ & $\chi^2/\rm dof$\\
\hline\noalign{\smallskip}
PG  1322+659  & 8.29 & $0.47^{+0.10}_{-0.12}$ &  $-0.94^{+0.14}_{-0.06}$ & $0.388^{+0.019}_{-0.020}$ & $25.7^{+3.5}_{-4.7}$ & 265.3 & 2.8/9 \\
PG 1115+407   & 8.18 & $0.41^{+0.06}_{-0.03}$ &  $-0.55^{+0.15}_{-0.15}$ & $0.420^{+0.020}_{-0.005}$ & $10.0^{+7.3}_{-3.1}$ & 241.2 & 4.3/5 \\
4C 10.06   & 9.00 & $0.60^{+0.15}_{-0.10}$ &  $0.7^{+0.05}_{-0.05}$&$0.415^{+0.020}_{-0.017}$ & $39.4^{+2.2}_{-2.0}$ & 194.7 & 4.9/7 \\
4C 39.25   & 9.33 & $0.47^{+0.03}_{-0.02}$ &  $0.998^{+0.000}_{-0.004}$&$0.344^{+0.008}_{-0.011}$ & $42.5^{+0.9}_{-1.4}$ & 158.3 & 11.5/11  \\
OS 562   & 8.92 & $0.33$ &  $0.96^{+0.03}_{-0.04}$&$0.505^{+0.027}_{-0.020}$ & $13.0^{+7.2}_{-9.6}$ & 142.5 & 4.3/7  \\
\hline\noalign{\smallskip}
\end{tabular}
\end{center}
\tablecomments{0.86\textwidth}{The black hole mass $M$ is taken from
\cite{2012ApJS..201...38T}. $T_{e}$ is the maximum value of the electron temperature of the corona, which is calculated with the best-fitting values of free parameters $a_{\ast}$ and $\dot{m}$.}
\end{table}

We use the general relativistic accretion disk corona model to fit
the observed optical/UV spectra and the hard X-ray continuum spectra
in $2-10$~keV. There are four free parameters in our model
calculations, i.e., the mass accretion rate $\dot{m}$, the spin
parameter $a_{\ast}$, the power fraction $f$ dissipated in the corona, and
the viewing angle $\theta$ with respect to disk axis. In the spectral
fitting, we find that the fraction $f$ is mainly determined by the
X-ray spectral index, which is almost
independent of the values of other parameters. For the broad-line
type I quasars, the values of the inclination angle are tuned in the
range of $\theta\la 45^\circ$.

\begin{figure}
   \centering
   \includegraphics[width=11.0cm, angle=0]{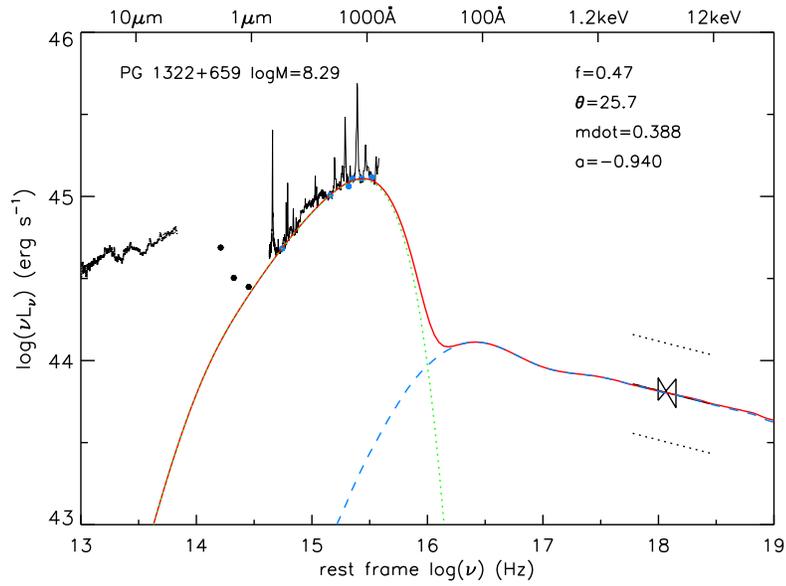}
\caption{The spectral fitting
results for PG~1322$+$659. The radio and near infrared data are not
used in our accretion disk corona spectral fittings. The photometric
points marked as blue filled circles are used for fitting. The solid black
line represents the power law X-ray spectrum in $2-10$~keV. The
red line represents the best fitting result with the black hole
mass $\rm log M=8.29$. The green dotted line represents the
spectrum of the cold disk, while the blue dashed line is for
the Compton radiation. The synchrotron and bremsstrahlung radiation which dominate
in radio bands are not plotted.}
\label{1322}
\end{figure}

\begin{figure}
   \centering
   \includegraphics[width=11.0cm, angle=0]{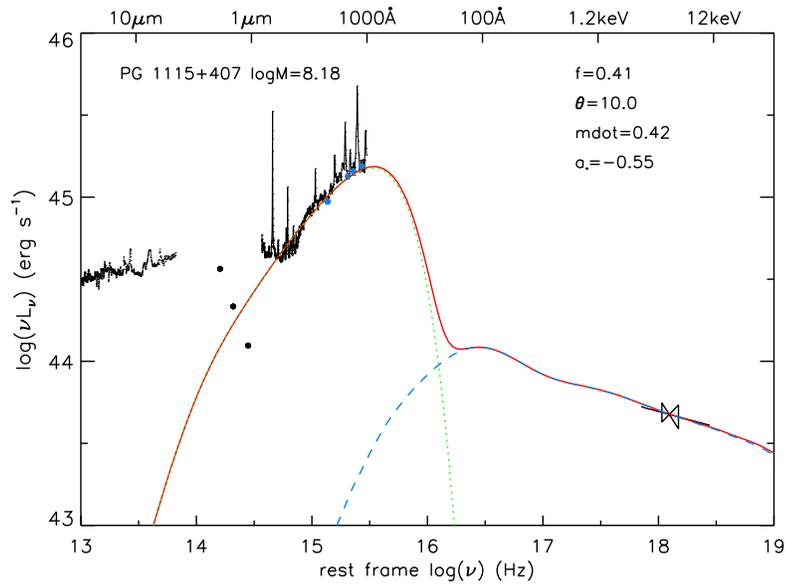} 
\caption{The best fitting result
for PG 1115$+$407.   \label{1115}}
\end{figure}

\begin{figure}
   \centering
   \includegraphics[width=11.0cm, angle=0]{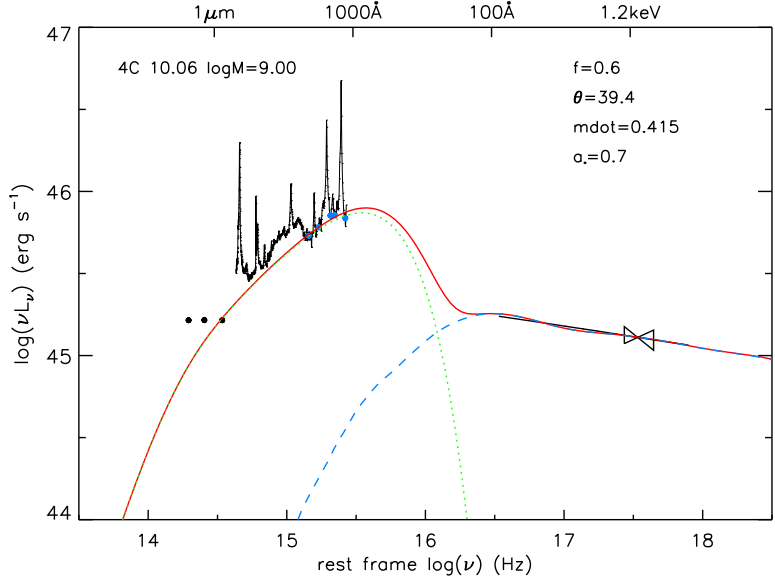}
\caption{The same as Figure
\ref{1322}, but for 4C~10.06.  \label{4c1006}}
\end{figure}

\begin{figure}
   \centering
   \includegraphics[width=11.0cm, angle=0]{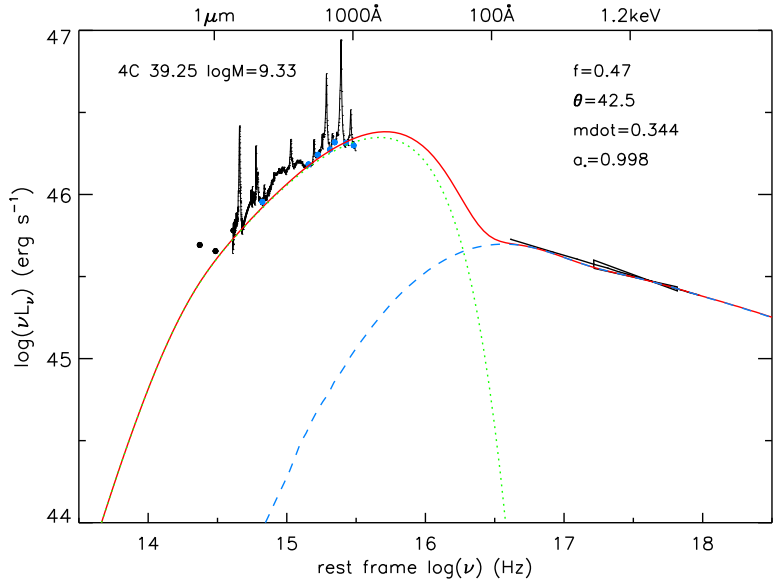}
\caption{The same as Figure
\ref{1322}, but for 4C~39.25. \label{4c3925}}
\end{figure}

\begin{figure}
   \centering
   \includegraphics[width=11.0cm, angle=0]{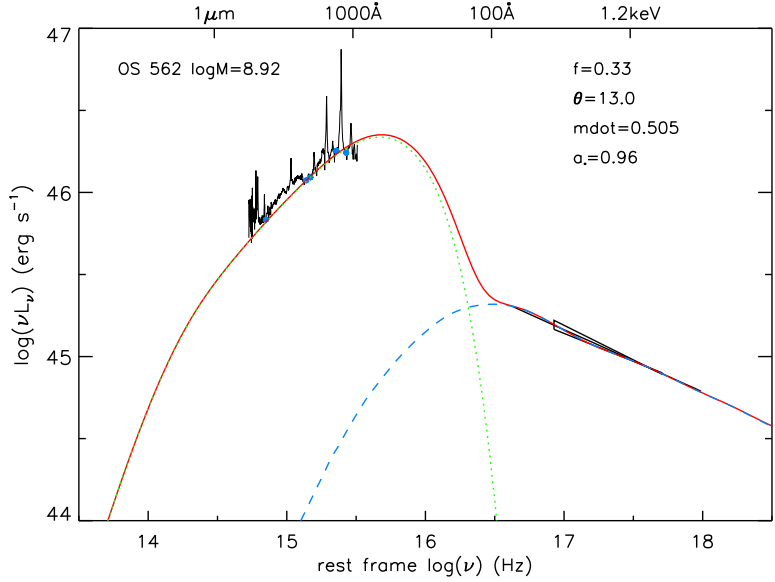}
\caption{The same as Figure
\ref{1322}, but for OS~562.  \label{os562}}
\end{figure}

We plot the spectral fitting results of the five quasars in Figures
\ref{1322}-\ref{os562}. In the optical/UV wavebands, the data points
(marked as blue filled circles) which are used in our spectral fitting are
extracted from the observed SED in the line-free continuum windows 
in order to avoid the contamination by emission lines, as done
in \cite{1997ApJ...475..469Z} and \cite{2001AJ....122..549V}.

\begin{figure}
   \centering
   \includegraphics[width=11.0cm, angle=0]{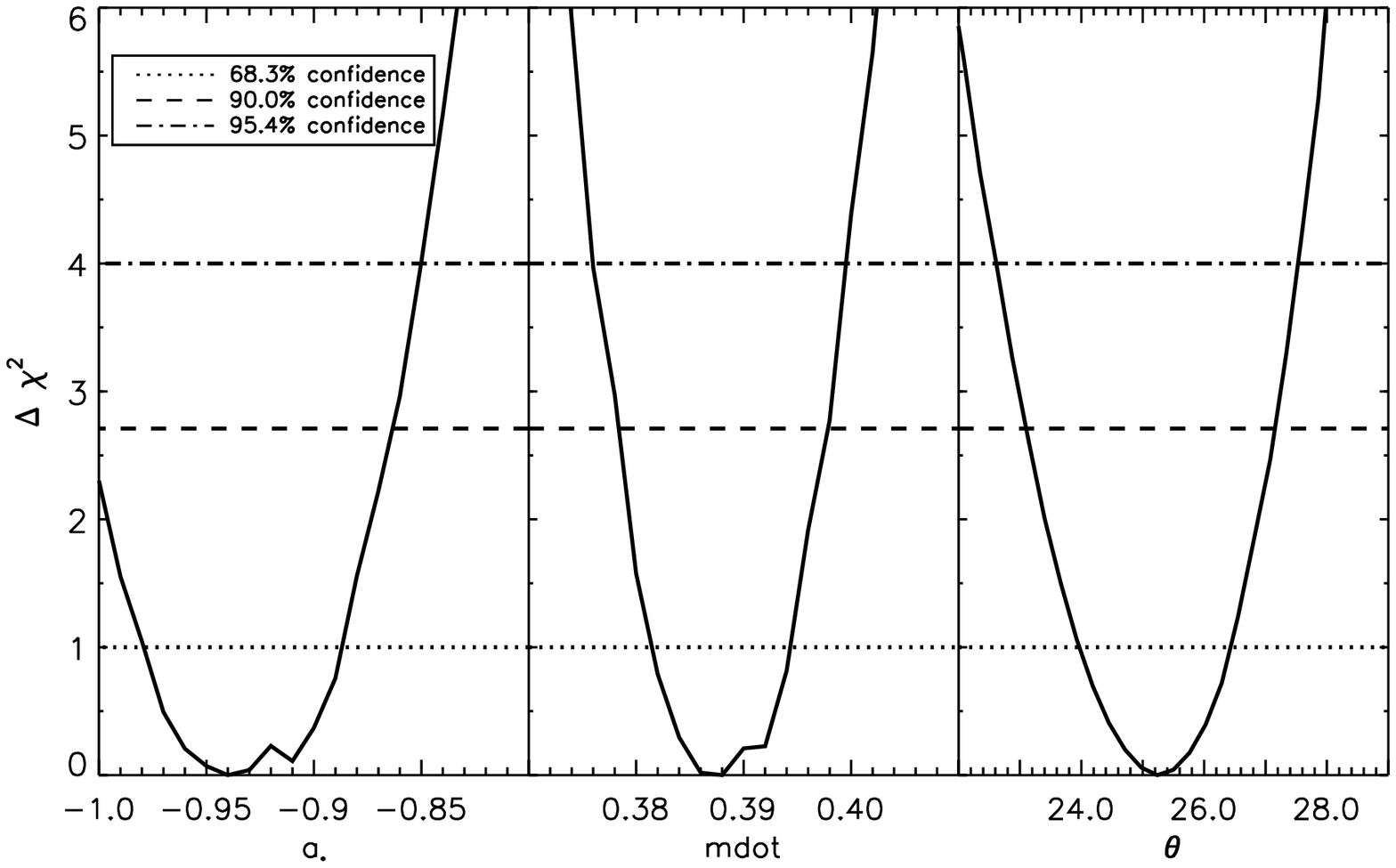}
\caption{Error contour for the spin parameter $a_{\ast}$, 
the mass accretion rate $\dot{m}$ and the inclination angle $\theta$. 
The three horizontal lines, from bottom to top, are corresponding
to the 68.3\%, 90.0\% and 95.4\% confidence, respectively.  \label{chi2}}
\end{figure}

In this work, the measured hard X-ray luminosity and the spectral
index are used for our spectral fitting. In the spectral fitting, the
errors of the fitting in the hard X-ray band (both the luminosity
and spectral index) are minimized by tuning the model parameters,
meanwhile the observed optical/UV spectrum is taken as upper limits.



The best-fitting results of the disk corona parameters are
listed in Table \ref{fitting}. The error of the parameter $f$ 
is determined according to the error of the hard X-ray spectral index $\alpha$ in Table 2, 
given that $f$ strongly depends on $\alpha$. The fitting error $\chi^{2}$ is
calculated in the parameter plane, and the distributions of
the error contour for the spin $a_{\ast}$, the mass accretion rate $\dot{m}$ 
and the inclination angle $\theta$ are plotted in Fig.~\ref{chi2}. 
The uncertainties on the three parameters correspond to 90\% confidence level.
. The 2D
confidence contours of $a_{\ast}$ and $\dot{m}$ for PG 1322$+$659 with the
measured black hole mass $M$ and fraction $f$ are plotted in Fig.~\ref{2d1322}.
We find that, the value of black hole spin parameter is constrained
in a narrow range $-1.0<a_{\ast}<-0.8$ with best-fitting value $a_{\ast}=-0.94$.
The best-fitting accretion rate $\dot{m}\simeq 0.4$, and the
inclination angle of the disk $\theta\simeq25^{\circ}$.
We know that uncertainties of the black hole spin constraints are
closely related to the errors of the black hole mass estimates. In
Fig.~\ref{m1322}, we plot the fitting results for PG 1322+659 with
the different values of $M$ reported in the literature, and find
that the constraints on the black hole spin parameter $a_{\ast}$ are
sensitive to the black hole mass estimate (see Table \ref{fm1322}).
Within the uncertainties of a factor of four on the mass measurements \citep{2006ApJ...641..689V},
using single epoch reverberation mapping method, from $2\times10^8M_{\odot}$ to $8\times10^8M_{\odot}$
with the step of $10^8M_{\odot}$, the best-fitting values of black hole spin
can vary from -0.94 to 0.998 (see Fig.~\ref{am}).
In Fig.~\ref{v1322}, we plot the fitting results for PG 1322+659
with a small shift $\Delta\log L_{\rm X}$ on the observed X-ray
luminosity to assess how the black hole spin $a_{\ast}$ is affected by
X-ray variability (see Table \ref{fm1322}).

We fit the emission line properties of PG 1322+659 by
subtracting the best spectral fitting continuum in Fig.~\ref{residual}, 
to compare with the typical broad line region
spectrum. The derived EW and FWHM for \mgii, \hb, and \hg \ are
consistent with those of the typical BLR spectrum given in
\citet{2011ApJS..194...45S}. In Fig.~\ref{height}, we plot the
relative disk thickness $H_{\rm d}/R$ varying with the mass
accretion rate $\dot{m}$. It is found that the thin disk 
approximation $H_{\rm d}/R\la 0.1$ can be achieved even if $\dot{m}>0.3$.
Note that when the mass 
accretion rate $\dot{m}$ is as high as $\sim 0.5$, 
$H_{\rm d}/R$ is only slightly larger than 0.1 for a typical value of $f\sim 0.5$.

\begin{table}
\begin{center}
\caption{The spectral fitting
results of PG 1322+659 \label{fm1322}}
\setlength{\tabcolsep}{0.2in} 
\begin{tabular}{cccccc}
\hline\noalign{\smallskip}
ID& log$M$&
$f$&$a_{\ast}$& $\dot{m}$& $\theta$\\ 
\hline\noalign{\smallskip}
1  & 8.29 & 0.47 &  -0.94 & 0.388& 25.7\\
\hline\noalign{\smallskip}
2  & 8.36 & 0.47& -0.75 & 0.316 & 29.8  \\
3  & 8.16 & 0.47& -0.94 & 0.520 & 0.0  \\
4  & 8.59 & 0.47& 0.90 & 0.06 & 30.0  \\
5  & 8.89 & 0.47& 0.94 & 0.03 & 43.0  \\
\hline\noalign{\smallskip}
6  & 8.29 & 0.47& -0.7 & 0.5 & 37.0  \\
7  & 8.29 & 0.47& -1.0 & 0.3 & 0.0  \\
\hline\noalign{\smallskip}
\end{tabular}
\end{center}
\tablecomments{0.86\textwidth}{The spectral fitting results with different black
hole masses in the literature (ID 2-5) and the variation of the
X-ray luminosity (ID 6,7). The fitting results are plotted in Fig.~\ref{m1322} and \ref{v1322} respectively.}
\end{table}

\section{Discussion}

The main features of quasar SED with a big blue bump in optical/UV
wavebands and a power-law hard X-ray continuum spectrum can be
reproduced by the accretion disk corona scenario (Galeev et al., 1979; Qiao \& Liu 2015).
There are various models which are developed in the exploring the radiation 
from the accretion disk and corona.
For instance, KERRBB is a relativistic multi-temperature blackbody model 
for the thin accretion disk around a Kerr black hole, which is usually used 
to fit the disk emission from disk-dominated X-ray binaries and 
quasars \citep{2005ApJS..157..335L}, while
SIMPL is a Comptonization model for modeling the power-law hard X-ray spectra of
X-ray binaries and AGNs \citep{2009PASP..121.1279S}.
The convolution between KERRBB and SIMPL in XSPEC \citep{1996ASPC..101...17A} 
could simulate the broad band SEDs of quasars,
namely, the optical BBB and power-law X-ray spectra, for the purpose of our paper here. In the resulting spectra of SIMPL*KERRBB, the hard X-ray power-law index depends on the parameter $\Gamma$, while the flux is determined by the parameters $f_{\rm SC}$ which is the fraction of the soft photons from the disk being scattered to high energy \citep[see Fig. 1 in][]{2009PASP..121.1279S}. As an alternative Comptonization model, a general relativistic model for an accretion disk corona surrounding a Kerr black hole was developed by \cite{2012ApJ...761..109Y}, in which the energy fraction dissipated into the corona was assumed in order to solve the pattern of the disk corona system. The emergent spectrum of the disk corona observed at infinity can be calculated when the values of the disk parameters are specified.

In this work, we try to fit the observed SEDs of quasars with the
general relativistic accretion disk corona model, in which the
values of the disk parameters for five quasars, especially the black
hole spin parameter $a_{\ast}$, are derived. The black hole masses of these
quasars are estimated with single epoch reverberation mapping method
\citep*[see][for the details]{2012ApJS..201...38T}.

The red tail of the BBB in optical wavebands is mostly emitted from
the outer region of the accretion disk, which can be used to
constrain the mass accretion rate $\dot{m}$ almost independent of
the black hole spin parameter $a_{\ast}$ if the black hole mass $M$
and the inclination angle $\theta$ are known
\citep[see][]{2011ApJ...728...98D}. In our accretion disk corona
model, the mass accretion rate $\dot{m}$ can be inferred similar to
their work. The peaks of the BBBs in most quasars are in the
far-UV/soft X-ray wavebands, which are usually unobserved. This is
sensitive to the black hole spin $a_{\ast}$, which is the main obstacle for
measuring the spin parameters $a_{\ast}$ of massive black holes with the
spectral fitting method. In the accretion disk corona model
considered in this work, the photons emitted from the disk,
including the unobserved far-UV/soft X-ray photons are inverse
Compton scattered in the hot corona, and the hard X-ray spectra
(luminosity and spectral shape) contain the information of the
incident spectra from the cold disk.

The fraction $f$ of the power dissipated in the corona to the total
power is a key parameter in the accretion disk-corona model. The
detailed physical processes for the energy transported from the disk
to the corona were explored by many different authors
\citep*[e.g.,][]{1998MNRAS.299L..15D,2000A&A...361..175M,
2002ApJ...572L.173L,2004ApJ...607L.107W,2009MNRAS.394..207C}. It is
found that both the hard X-ray spectral index and the hard X-ray
bolometric correction factor $L_{\rm bol}/L_{\rm X,2-10 keV}$ are
correlated with the Eddington ratio
(e.g.,
Wang et al., 2004; Shemmer et al., 2006; Vasudevan \& Fabian, 2007; 
Zhou \& Zhao 2010; Fanali et al., 2013),
which can be reproduced roughly by the accretion disk corona model
on the assumption that the corona is heated by the reconnection of
the magnetic fields generated by buoyancy instability in the cold
accretion disk \citep{2009MNRAS.394..207C}. However, the precise
value of $f$ is still undetermined in the theoretical model
calculations. In this work, we adopt the fraction $f$ as a free
parameter in the spectral fittings.

In the spectral fitting of a quasar SED with measured black
hole mass, the value of $f$ is dominantly determined by the hard
X-ray photon index $\Gamma$, which is almost insensitive to all
other model parameters
\citep*{2009MNRAS.394..207C,2012ApJ...761..109Y}. With the derived
value of $f$ from $\Gamma$, the mass accretion rate $\dot{m}$ and
the inclination angle $\theta$ are constrained with the luminosity and
spectral shape in optical waveband (the red tail of the BBB). These
two parameters are almost independent of the black hole spin
parameter $a_{\ast}$, because the emission in this waveband is dominantly
from the outer region of the disk \citep*[][]{2011ApJ...728...98D} .
Finally, the black hole spin parameter $a_{\ast}$ is constrained mainly
with the observed hard X-ray luminosity.

In Figures \ref{1322}-\ref{os562}, we plot the fitting results of
five quasars. The values of black hole spin parameter $a_{\ast}$ are well
constrained with small uncertainties, which are in the range of
$\sim -0.94$ to $0.998$ (see Table 3). The estimated accretion rate
is moderate, approximately 0.4 except OS 562 with
$\dot{m}\simeq0.5$, and these quasars are possibly viewed with small
inclination angles $<45^{\circ}$, which support the classification
of the sources as type 1 quasars. The fractions $f$ of the power
radiated in the corona to the total dissipated power in the disk are
$\sim 0.3-0.6$ (see Table \ref{fitting}).

\begin{figure}
   \centering
   \includegraphics[width=11.0cm, angle=0]{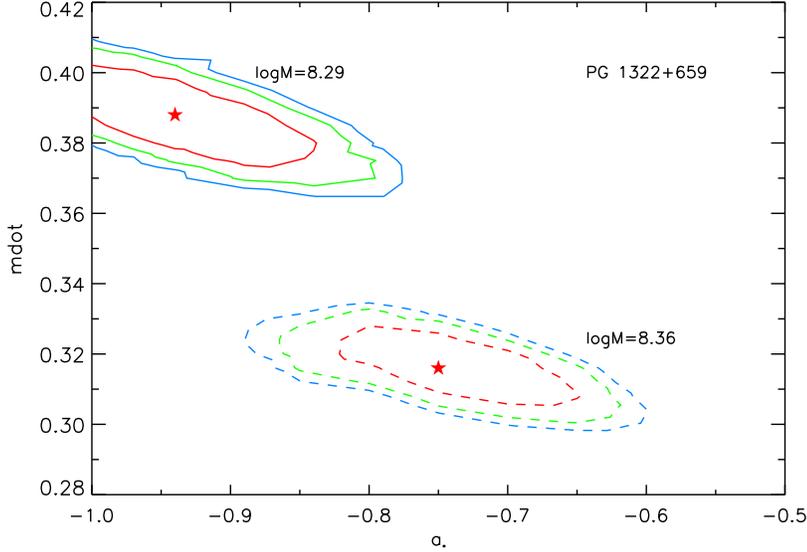}
\caption{The confidence contours
(red:68.3\%, green:90.0\%, blue:95.4\%) for the black hole spin $a_{\ast}$ and the accretion
rate $\dot{\rm m}$ of PG 1322+659, for two black hole masses $\rm log M=8.36$ and$\rm log M=8.29$.
For $\rm log M=8.16$, the spectrum is not fitted well. The star indicates the position
of the best fit. The fraction $f=0.47$ is adopted.
\label{2d1322}}
\end{figure}

In the spectral fitting on the SEDs of these sources in this work, the
electron temperatures of the coronas are in the range of
$100-300$~keV, which are roughly consistent with the previous works
\citep*[e.g.,][]{1996A&AS..120C.553Z,2002ApJ...572L.173L}.
The maximal temperatures of the electrons in the coronas for
the sources are listed in Table 3. The corresponding cutoff energy
of the X-ray spectra is around several hundred keV
\citep*[see][]{2012ApJ...761..109Y}. No direct observations in this
energy band are available for these five sources. The cutoff energy
in the power law X-ray spectra is observed only in a small fraction
of AGNs, which is typically in the range of $30-300$~keV
\citep*[][]{2013MNRAS.433.1687M}. This is consistent with our
results.

We plot the distribution of the fitting error $\Delta \chi^2$ for
the spin $a_{\ast}$, the mass accretion rate $\dot{m}$ and the inclination angle $\theta$ in Fig.~\ref{chi2},
in which the uncertainties of the fitted parameters are given by $90 \%$ confidence levels.
We also plot 2D confidence contours in Fig.~\ref{2d1322}, and find that the black hole spin $a_{\ast}$ is indeed
well constrained at fairly good accuracy if the black hole mass is
accurately measured.


The observed hard X-ray spectra may be affected by the
reflection components (Gou et al., 2011,
2014). However, there are observational evidences
that the reflection components are absent in many bright quasars
(high luminosities)
\citep*[][]{1997ApJ...488L..91N,2005MNRAS.364..195P,2007A&A...463...79G}.
The \textit{XMM-Newton} observations of a sample of high redshift
quasars show that the hard X-ray spectra ($>10$~keV in the rest
frame of the sources) can be well fitted for each source by a simple
power law, and no Compton reflection humps are present
\citep*[][]{2005MNRAS.364..195P}. The five sources in this work are
all luminous quasars with $\dot{m}\ga 0.3$, and therefore we believe
that their hard X-ray continuum spectra may probably not be
contaminated by the reflection components. Further X-ray observations in high energy with/without Compton reflection hump for these sources may help resolve this issue.

\begin{figure}
   \centering
   \includegraphics[width=11.0cm, angle=0]{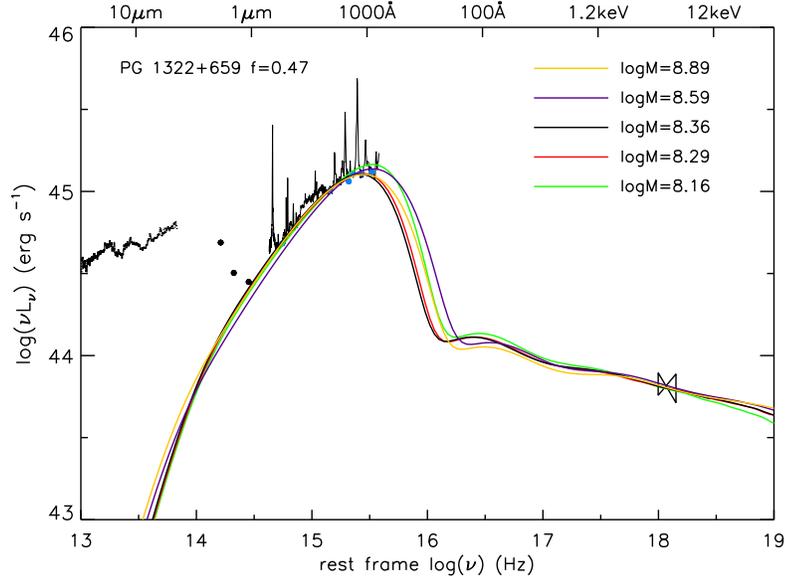}
\caption{The same as Fig.~\ref{1322},
but with different black hole masses reported in the literature:
$\rm log M=8.36$ \citep{2006ApJ...641..689V}; $\rm log M=8.16$
\citep{2007ApJ...661..660K} as well as increased by a factor of two
and four:$\rm log M=8.59$, $\rm log M=8.89$. The fitting results
are, yellow line: $a_{\ast}=0.94$, $\dot{m}=0.03$, $\theta=43.0$; purple
line: $a_{\ast}=0.90$, $\dot{m}=0.066$, $\theta=30.0$; black line:
$a_{\ast}=-0.75$, $\dot{m}=0.316$, $\theta=29.8$; red line: $a_{\ast}=-0.94$,
$\dot{m}=0.388$, $\theta=25.7$; green line: $a_{\ast}=-0.94$,
$\dot{m}=0.52$, $\theta=0.0$. \label{m1322}}
\end{figure}

\begin{figure}
   \centering
   \includegraphics[width=11.0cm, angle=0]{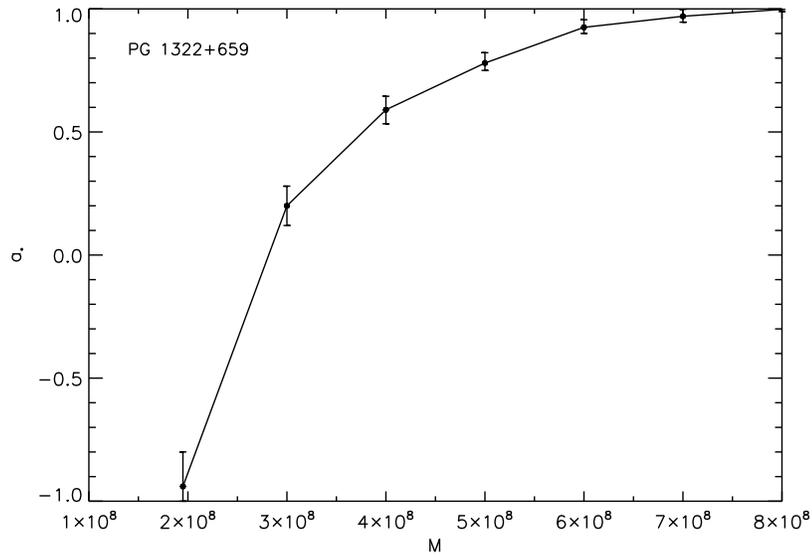}
\caption{The best-fitting values
of black hole spin for the different black hole masses from  $2\times10^8M_{\odot}$ to $8\times10^8M_{\odot}$,
with the step of $10^8M_{\odot}$.
\label{am}}
\end{figure}

Another uncertainty of the black hole spin constraints is caused by
the errors of the black hole mass estimates, because the constraints
on the spin parameter $a_{\ast}$ is sensitive to the black hole mass (see
Table \ref{fm1322}). The black hole spin of quasar SDSS
J094533.99+100950.1 is derived by fitting its multi-band
(IR/optical/UV) continuum spectrum. The black hole spin $a_{\ast}$ varies
from $-1.0$ to $0.998$ if the black hole mass is taken as a free
parameter, which means that the black hole mass is crucial in
constraining the black hole spin $a_{\ast}$
\citep*[][]{2011MNRAS.415.2942C}. In order to evaluate how the
constraint on the spin parameter $a_{\ast}$ is affected by the black hole
mass, we re-fit the SED of PG 1322+659 with different values of
black hole mass given in different works (see Table \ref{fm1322}).
For the black hole mass $\log M=8.36$ estimated with \civ \ line
\citep{2006ApJ...641..689V}, the best fitting of black-hole spin
appears as $a_{\ast}=-0.75$, while $a_{\ast}=-0.94$ if $\log M=8.29$ is adopted.
For the black hole mass $\log M=8.16$ estimated by
\citet{2007ApJ...661..660K}, no satisfactory spectral fitting is
available.
In most previous works, the uncertainty of the black hole
mass estimates with single epoch reverberation mapping approach is
within a factor of four
\citep*[e.g.,][]{2004A&A...424..793W,2004ApJ...613..682P,2006ApJ...641..689V}.
Increasing the mass $\log M=8.29$ by a factor of two and four, i.e.,
$\log M=8.59, 8.89$, we find that the black hole spin parameters are
constrained as $a_{\ast}=0.9, 0.94$ respectively (see Fig.~\ref{m1322}).
With typical accuracy of the black hole mass measurements, the best
fitted black hole spins are in a large range from -0.94 to 0.998
(see Fig.~\ref{am}), which indicates that no tight constraint on
the black hole spin is achieved with the uncertainty of a factor of
four in mass estimates. However, it is not unique.
Actually this issue is already discovered by previous works when the continuum-fitting method is applied to both AGNs and X-ray binaries. For instance, as for AGNs, the best-fitting values of the spin a for different black hole masses and the inclination angles can be found in Fig 5 of \cite{2013MNRAS.434.1955D}, which showed that the spin could vary from -1 to 1 if the mass is in a large range; as for X-ray binaries, even the system parameters of LMC X-3 are fairly well determined, the spin can spread over a large range with different combinations of the blacho hole mass and inclination angle (see Table 2 of Kubota et al., 2010). The improved estimate of the spin of LMC X-3 $a_{\ast}=0.25$ \citep{2014ApJ...793L..29S}, with the dynamical parameters to date \citep{2014ApJ...794..154O}. 
Although the spin covers almost all the possible range for the spin parameter due to large errors of the black hole mass, fitting the disk-corona model to the optical/UV and X-ray SED in our paper is meriting exploration.

The main uncertainty of traditional
reverberation mapping analysis originates from the virial
coefficient or normalizing factor. Recently,
\citet{2011ApJ...730..139P} proposed a method to analyze
reverberation mapping data, with which the geometry and kinematics
of the BLR can be well constrained by modeling the continuum light
curve and broad line profiles directly with the velocity resolved
broad line data. This allows to make a measurement of the black hole
mass that does not depend on the virial coefficient
\citep*[][]{2011ApJ...730..139P,2012ApJ...754...49P}. The accuracy
of the black hole mass estimates with this method has been
significantly improved
\citep*{2011ApJ...743L...4B,2012ApJ...754...49P,2013ApJ...779..110L}.
The preliminary results show that five black hole masses have been
measured at an uncertainty of $\sim 0.2-0.4$ dex depending upon data
quality with this method \citep*[][]{2014MNRAS.445.3073P}. We search
the literature, and find that the X-ray spectral data are available
only for the two sources. Unfortunately, the X-ray photon spectral
index $\Gamma <2$ for these two sources, which are unsuitable for
our present investigation. The black hole spins can be well
constrained with the method in this work, when a variety of black
hole masses are measured at an uncertainty of $\sim 0.2$ dex with
the velocity resolved reverberation mapping method in the near
future.

In our model calculations, the hard X-ray emission of these
quasars is assumed to originate predominantly from the corona above
the disk, which is true for radio quiet sources. However, it should
be cautious for radio loud quasars, of which the observed hard X-ray
emission may be contaminated by the beamed jet emission
\citep[e.g.,][]{2012ApJ...748..119C}. We find that OS 562 is a Flat
Spectrum Radio Quasar (FSRQ), which is also characterized as
low-spectral peaked source with the synchrotron peak frequency
$\nu_{\rm p}<10^{\rm 14} \rm Hz$ \citep{2009AJ....137.3718L}. The
Very Long Baseline Array (VLBA) observations indicate that this
source is observed with a small inclination angle with respect to
the jet axis \citep{2001ApJ...562..208L}, which is consistent with
the viewing angle $\theta=13^{\circ}$ derived in our spectral fitting
(see Table \ref{fitting}). \cite{2001A&A...375..739D} analyzed the
X-ray spectral of 268 blazars including BL Lacs and FSRQs, and found
that the X-ray spectra of FSRQs are always hard, i.e., the average
X-ray spectral index $\alpha\simeq -0.76$ ($\Gamma\simeq 1.76$). The
X-ray spectral index of OS~562 is $-1.38$, much softer than a
typical FSRQ, which may imply that the X-ray emission is
predominantly from the corona. This is the case also for the other
two radio quasars. The radio core dominance $R$ (defined as
$R=L_{\rm core}/L_{\rm ext}$) is usually taken as a radio
orientation indicator
\citep{1993ApJ...407...65G,1995ApJ...448L..81W}. The radio core
dominance parameters of the other two radio loud quasars in this
work are, $\log R=-0.658$ (4C 10.06) and $-2.22$ (4C 39.25)
\citep*{2013MNRAS.435.3251R}, which imply that the jets are inclined
at large angles with respect to the line of sight. This is
consistent with our fitting results of $\theta\simeq40^{\circ}$ for
these two sources. The beaming effects for the jets viewed at such
large angles are always unimportant
\citep*[e.g.,][]{2013ApJ...770...31W}. This strengthens the
conclusion that the beamed X-ray emission from the jets can be
neglected compared with that from the coronas in these two sources.

\begin{figure}
   \centering
   \includegraphics[width=11.0cm, angle=0]{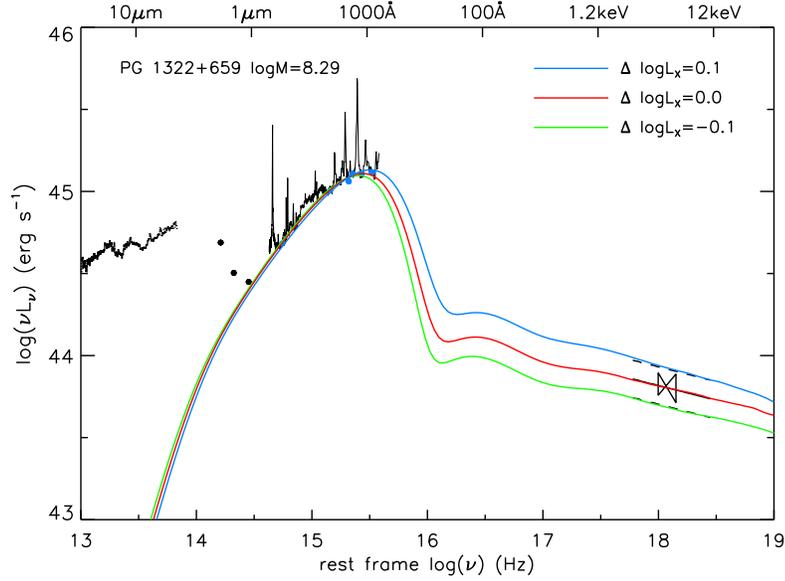}
\caption{The same as Fig.~\ref{1322},
but with a small shift $\rm \Delta logL_{\rm X}=\pm0.1$ on the
observed X-ray luminosity (dashed lines) to assess how the black
hole spin $a_{\ast}$ is affected by X-ray variability. The parameters
$\rm logM=8.29$ and $f=0.47$ are adopted. The fitting results are,
blue line: $a_{\ast}=-0.7$, $\dot{m}=0.5$, $\theta=37.0$;
red line: $a_{\ast}=-0.94$, $\dot{m}=0.388$, $\theta=25.7$;
green line: $a_{\ast}=-1.0$, $\dot{m}=0.3$, $\theta=0.0$. \label{v1322}}
\end{figure}

\begin{figure}
   \centering
   \includegraphics[width=11.0cm, angle=0]{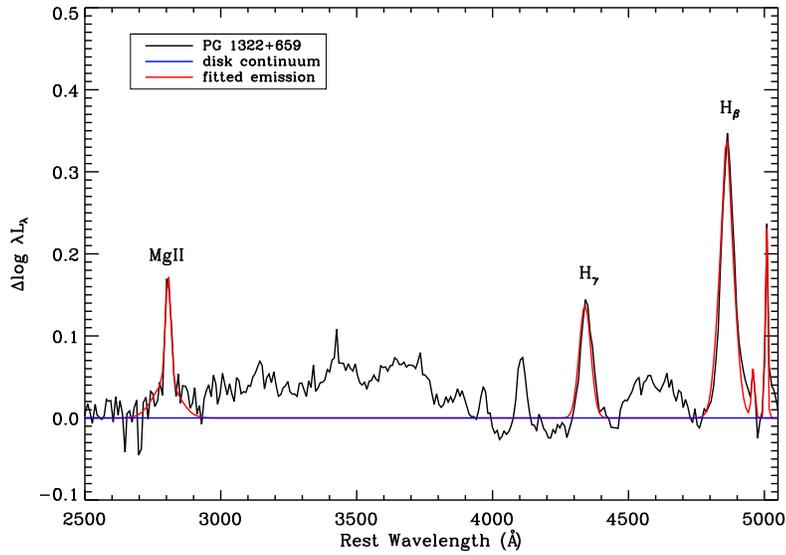}
\caption{The residual emission
line spectrum for PG 1322+659 marked as black line, with the best
fit continuum from the Fig.~\ref{1322} being subtracted. The fitted
emission lines of \mgii,\hg and \hb are marked with red lines.
\label{residual}}
\end{figure}

\begin{figure}
   \centering
   \includegraphics[width=11.0cm, angle=0]{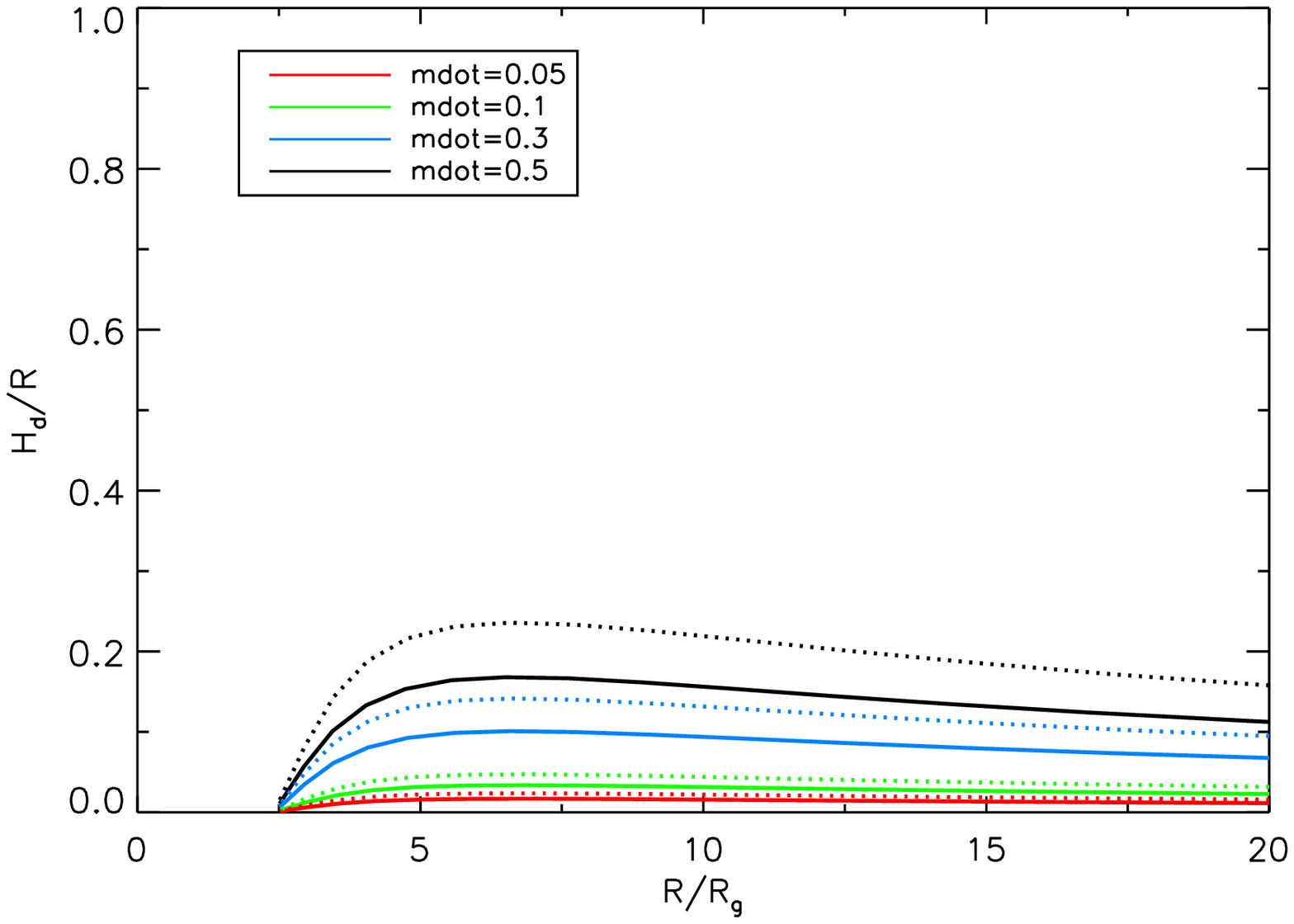} 
\caption{The relative disk
thickness $H_{\rm d}/R$ as functions of radius with different
accretion rates $\dot{m}$ for a given black hole mass $M=10^8M_{\odot}$
and spin $a_{\ast}=0.9$. The solid lines represent the
results for disks with coronas, while the dotted lines are for bare
accretion disks without corona. The colored lines represent the
model calculations with different values of accretion rates.
\label{height}}
\end{figure}

Although there are no additional information of X-ray variability
for 4C 10.06 and OS 562, some previous observations indicated that
the other three sources have low X-ray variability,
$\Delta\log L_{\rm X}=0.03, 0.1$ for PG 1322+659 and PG 1115+407 \citep{2001A&A...367..470G},
$10\%\sim30\%$ variation for a sample of flat-spectrum radio quasars including 4C 39.25.
In Fig.~\ref{v1322}, we show that the constraint on the
black hole spin $a_{\ast}$ is affected by the X-ray variability. It is
found that the spin parameter becomes $a_{\ast}=-1$ for $\Delta\log L_{\rm
X}=-0.1$, while $a_{\ast}=-0.7$ for $\Delta\log L_{\rm X}=0.1$. This
implies that the sources with violent variability are not suitable
for present investigation on black hole spin. Our model is
applicable for steady accretion disk corona systems. For those
slightly variable sources, the mean observed continuum spectrum can
still be used to constrain the black hole spin parameters.

Although the soft X-ray spectral data are not used in our model 
fitting on the SEDs, given the origin of the soft X-ray excess observed in AGNs 
is still controversial \citep{2012MNRAS.420.1848D}, it is
still necessary to investgate the influence of the soft X-ray excess on determining
black hole spin. It was done for MCG -06-30-15 which showed a soft
excess below $\sim$ 0.7 keV \citep{2006ApJ...652.1028B}. Fitting physical
models to the 
broadband spectrum at 2-10 keV with broad iron line,
the spin of the black hole in MCG -06-30-15 is well constrained to be $a_{\ast}=0.997$.
Moreover, if including the observed features at the lower energy bands 0.6-2 keV 
and introducing additional components as the possible origin of soft X-ray
excess, the best fits to the overall spectrum estimated the black hole 
spin of $a_{\ast}=0.997$ or 0.989, which indicated the weak influence of the soft X-ray
excess on determining the spin of the black hole. Given that we don't
have the observation of soft X-ray excess for the five sources
in our work, we try to discuss the influence of the soft X-ray excess 
in terms of power dissipation. We assume that 20\% of the dissipated disk power
is radiated as soft X-ray excess component \citep{2014ApJ...787...83M}, then the parameter $f$ is defined
as the ratio of the dissipated power in the corona to the remaining power 
in the disk, i.e., $f=Q_{\rm cor}^{+}/(80\%Q_{\rm dissi}^{+})$.
We re-fit the SED of PG 1322+659 with this definition of $f$, and
find that the best-fitting value of black hole spin $a_{\ast}=-0.9$, showing small
difference in the spin parameter. The remaining model parameters are $f=0.58$, $\dot{m}=0.378$, 
and $\theta=24.7$, respectively.

For the traditional thin accretion disk model, the mass accretion
rate $\dot{m}\la 0.3$ is required in order to satisfy the
geometrically thin approximation, i.e., $H_{\rm d}/R\la 0.1$
\citep*[e.g.,][]{1989MNRAS.238..897L,1997ApJ...489..865E,2012MmSAI..83..466K,2013LRR....16....1A}.
We find that $\dot{m}\ga 0.3$ is required for the spectral fitting in
five sources in this work. Our calculations show that the situation
is different for the accretion disk with corona, because a fraction
of gravitational power is transported into the corona, which reduces
the radiation pressure of the disk for a given mass accretion rate
$\dot{m}$. In Fig.~\ref{height}, we find that the geometrically
thin disk assumption is still valid even if $0.3<\dot{m}<0.5$ when a
typical value of $f=0.5$ is adopted. This implies that the thin
accretion disk corona model is applicable for all quasars in this
work. For more luminous quasars with $\dot{m}\ga 0.5$, a general
relativistic slim accretion disk corona model is required, which
will be reported in our future work.

In this work, we tentatively propose a variant of the continuum-fitting method to constrain the black
hole spin parameter $a_{\ast}$ in AGNs with a general relativistic accretion corona
model. A reliable way to test the feasibility of this proposed method is to apply it to a few stellar-mass black hole X-ray binaries with available estimates of the black hole spin from both continuum-fitting and broad Fe K${\alpha}$ line fitting methods, to see whether the similar results can be obtained. In this paper, it is found that the black hole spin parameter $a_{\ast}$ can be
well constrained, if the hard X-ray continuum spectrum and the black
hole mass are accurately measured. We have not studied the
statistics of the black hole spin parameter $a_{\ast}$ in this work, due to
the small number of the quasars, which can provide useful
information when a large sample of quasars with measured black hole
spins $a_{\ast}$.

\begin{acknowledgements}
We thank Zhaohui Shang for the SED data, careful reading the
manuscript, and helpful comments/suggestions. Weimin Yuan, Tinggui
Wang, Minfeng Gu, Shiying Shen, and Jianmin Wang are thanked for the
helpful discussion. Hengxiao Guo is thanked for fitting the emission
lines in Fig.~\ref{residual}. This work is supported by the NSFC
(grants 11173043, 11121062, 11233006, 11073020, 11373056, and 11473054), the Fundamental
Research Funds for the Central Universities (WK2030220004), the
CAS/SAFEA International Partnership Program for Creative Research
Teams (KJCX2-YW-T23), and Shanghai Municipality.
\end{acknowledgements}




\begin{thebibliography}
\small \setlength{\itemindent}{-3mm} \setlength{\itemsep}{-0.5mm}
\setlength{\baselineskip}{4.5mm}

\bibitem[Abramowicz
\& Fragile(2013)]{2013LRR....16....1A} Abramowicz, M.~A., \& Fragile, P.~C.\ 2013, Living Reviews in Relativity, 16, 1


\bibitem[Alexander
\& Hickox(2012)]{2012NewAR..56...93A} Alexander, D.~M., \& Hickox, R.~C.\ 2012, Nature, 56, 93

\bibitem[Antonucci(1999)]{1999ASPC..161..193A} Antonucci, R.\ 1999, High
Energy Processes in Accreting Black Holes, 161, 193

\bibitem[Arnaud(1996)]{1996ASPC..101...17A} Arnaud, K.~A.\ 1996, 
Astronomical Data Analysis Software and Systems V, 101, 17 

\bibitem[Barr et al.(1985)]{1985MNRAS.216P..65B} Barr, P., White, N.~E.,
\& Page, C.~G.\ 1985, \mnras, 216, 65P

\bibitem[Barth et al.(2011)]{2011ApJ...743L...4B} Barth, A.~J., Pancoast,
A., Thorman, S.~J., et al.\ 2011, \apjl, 743, L4


\bibitem[Bechtold et al.(1987)]{1987ApJ...314..699B} Bechtold, J., Czerny,
B., Elvis, M., Fabbiano, G., \& Green, R.~F.\ 1987, \apj, 314, 699

\bibitem[Beloborodov(1999)]{1999ApJ...510L.123B} Beloborodov, A.~M.\ 1999,
\apjl, 510, L123

\bibitem[Bentz et al.(2009)]{2009ApJ...705..199B} Bentz, M.~C., Walsh,
J.~L., Barth, A.~J., et al.\ 2009, \apj, 705, 199

\bibitem[Blaes(2013)]{2013SSRv..tmp...58B} Blaes, O.\ 2013, \ssr, 58


\bibitem[Blandford
\& Znajek(1977)]{1977MNRAS.179..433B} Blandford, R.~D., \& Znajek, R.~L.\ 1977, \mnras, 179, 433


\bibitem[Blandford
\& Payne(1982)]{1982MNRAS.199..883B} Blandford, R.~D., \& Payne, D.~G.\ 1982, \mnras, 199, 883

\bibitem[Berti
\& Volonteri(2008)]{2008ApJ...684..822B} Berti, E., \& Volonteri, M.\ 2008, \apj, 684, 822


\bibitem[Brenneman
\& Reynolds(2006)]{2006ApJ...652.1028B} Brenneman, L.~W., \& Reynolds, C.~S.\ 2006, \apj, 652, 1028


\bibitem[Brinkmann et
al.(1997)]{1997A&A...319..413B} Brinkmann, W., Yuan, W., \& Siebert, J.\ 1997, \aap, 319, 413


\bibitem[Calderone et al.(2013)]{2013MNRAS.431..210C} Calderone, G.,
Ghisellini, G., Colpi, M., \& Dotti, M.\ 2013, \mnras, 431, 210




\bibitem[Cao(2009)]{2009MNRAS.394..207C} Cao, X.\ 2009, \mnras, 394, 207



\bibitem[Chandrasekhar(1960)]{1960ratr.book.....C} Chandrasekhar, S.\ 1960,
New York: Dover, 1960,


\bibitem[Chen et al.(2012)]{2012ApJ...748..119C} Chen, L., Cao, X.,
\& Bai, J.~M.\ 2012, \apj, 748, 119



\bibitem[Chiang(2002)]{2002ApJ...572...79C} Chiang, J.\ 2002, \apj, 572, 79

\bibitem[Cunningham(1975)]{1975ApJ...202..788C} Cunningham, C.~T.\ 1975,
\apj, 202, 788


\bibitem[Czerny et al.(2011)]{2011MNRAS.415.2942C} Czerny, B., Hryniewicz,
K., Niko{\l}ajuk, M., \& S{\c a}dowski, A.\ 2011, \mnras, 415, 2942



\bibitem[Davis
\& Laor(2011)]{2011ApJ...728...98D} Davis, S.~W., \& Laor, A.\ 2011, \apj, 728, 98


\bibitem[de La Calle P{\'e}rez et
al.(2010)]{2010A&A...524A..50D} de La Calle P{\'e}rez, I., Longinotti, A.~L., Guainazzi, M., et al.\ 2010, \aap, 524, A50

\bibitem[Di Matteo(1998)]{1998MNRAS.299L..15D} Di Matteo, T.\ 1998, \mnras,
299, L15

\bibitem[Donato et
al.(2001)]{2001A&A...375..739D} Donato, D., Ghisellini, G., Tagliaferri, G., \& Fossati, G.\ 2001, \aap, 375, 739

\bibitem[Done et al.(2012)]{2012MNRAS.420.1848D} Done, C., Davis, S.~W.,
Jin, C., Blaes, O., \& Ward, M.\ 2012, \mnras, 420, 1848

\bibitem[Done et al.(2013)]{2013MNRAS.434.1955D} Done, C., Jin, C., 
Middleton, M., \& Ward, M.\ 2013, \mnras, 434, 1955

\bibitem[Dotti et al.(2013)]{2013ApJ...762...68D} Dotti, M., Colpi, M.,
Pallini, S., Perego, A., \& Volonteri, M.\ 2013, \apj, 762, 68



\bibitem[Esin et al.(1997)]{1997ApJ...489..865E} Esin, A.~A., McClintock,
J.~E., \& Narayan, R.\ 1997, \apj, 489, 865

\bibitem[Fanali et al.(2013)]{2013MNRAS.433..648F} Fanali, R., Caccianiga, 
A., Severgnini, P., et al.\ 2013, \mnras, 433, 648 


\bibitem[Ferrarese
\& Merritt(2000)]{2000ApJ...539L...9F} Ferrarese, L., \& Merritt, D.\ 2000, \apjl, 539, L9



\bibitem[Frank et al.(1992)]{1992apa..book.....F} Frank, J., King, A.,
\& Raine, D.\ 1992, Camb.~Astrophys.~Ser., Vol.~21,,


\bibitem[Galeev et al.(1979)]{1979ApJ...229..318G} Galeev, A.~A., Rosner,
R., \& Vaiana, G.~S.\ 1979, \apj, 229, 318

\bibitem[Ghisellini et al.(1993)]{1993ApJ...407...65G} Ghisellini, G.,
Padovani, P., Celotti, A., \& Maraschi, L.\ 1993, \apj, 407, 65


\bibitem[Gilli et
al.(2007)]{2007A&A...463...79G} Gilli, R., Comastri, A., \&
Hasinger, G.\ 2007, \aap, 463, 79

\bibitem[Goosmann et
al.(2006)]{2006A&A...454..741G} Goosmann, R.~W., Czerny, B., Mouchet, M., et al.\ 2006, \aap, 454, 741


\bibitem[Gou et al.(2011)]{2011ApJ...742...85G} Gou, L., McClintock, J.~E.,
Reid, M.~J., et al.\ 2011, \apj, 742, 85

\bibitem[Gou et al.(2014)]{2014ApJ...790...29G} Gou, L., McClintock, J.~E., 
Remillard, R.~A., et al.\ 2014, \apj, 790, 29 

\bibitem[Grupe et
al.(2001)]{2001A&A...367..470G} Grupe, D., Thomas, H.-C., \& Beuermann, K.\ 2001, \aap, 367, 470

\bibitem[Haardt
\& Maraschi(1991)]{1991ApJ...380L..51H} Haardt, F., \& Maraschi, L.\ 1991, \apjl, 380, L51

\bibitem[Haardt et al.(1994)]{1994ApJ...432L..95H} Haardt, F., Maraschi,
L., \& Ghisellini, G.\ 1994, \apjl, 432, L95


\bibitem[Hirose et al.(2009)]{2009ApJ...704..781H} Hirose, S., Blaes, O.,
\& Krolik, J.~H.\ 2009, \apj, 704, 781



\bibitem[Hopkins et al.(2008)]{2008ApJS..175..356H} Hopkins, P.~F.,
Hernquist, L., Cox, T.~J., \& Kere{\v s}, D.\ 2008, \apjs, 175, 356



\bibitem[Hubeny et al.(2001)]{2001ApJ...559..680H} Hubeny, I., Blaes, O.,
Krolik, J.~H., \& Agol, E.\ 2001, \apj, 559, 680




\bibitem[Kaspi et al.(2000)]{2000ApJ...533..631K} Kaspi, S., Smith, P.~S.,
Netzer, H., et al.\ 2000, \apj, 533, 631


\bibitem[Kawakatu et al.(2007)]{2007ApJ...661..660K} Kawakatu, N.,
Imanishi, M., \& Nagao, T.\ 2007, \apj, 661, 660


\bibitem[King(2012)]{2012MmSAI..83..466K} King, A.\ 2012, \memsai, 83, 466

\bibitem[Kishimoto et al.(2008)]{2008Natur.454..492K} Kishimoto, M.,
Antonucci, R., Blaes, O., et al.\ 2008, \nat, 454, 492

\bibitem[Koratkar
\& Blaes(1999)]{1999PASP..111....1K} Koratkar, A., \& Blaes, O.\ 1999, \pasp, 111, 1

\bibitem[Kormendy
\& Richstone(1995)]{1995ARA&A..33..581K} Kormendy, J., \& Richstone, D.\ 1995, \araa, 33, 581


\bibitem[Kubota et al.(2010)]{2010ApJ...714..860K} Kubota, A., Done, C., 
Davis, S.~W., et al.\ 2010, \apj, 714, 860


\bibitem[Laor
\& Netzer(1989)]{1989MNRAS.238..897L} Laor, A., \& Netzer, H.\ 1989, \mnras, 238, 897


\bibitem[Laor(1991)]{1991ApJ...376...90L} Laor, A.\ 1991, \apj, 376, 90

\bibitem[Laor
\& Davis(2014)]{2014MNRAS.tmp...60L} Laor, A., \& Davis, S.~W.\ 2014, \mnras, 60


\bibitem[Li et al.(2005)]{2005ApJS..157..335L} Li, L.-X., Zimmerman, E.~R.,
Narayan, R., \& McClintock, J.~E.\ 2005, \apjs, 157, 335

\bibitem[Li et al.(2013)]{2013ApJ...779..110L} Li, Y.-R., Wang, J.-M., Ho,
L.~C., Du, P., \& Bai, J.-M.\ 2013, \apj, 779, 110



\bibitem[Li et al.(2009)]{2009ApJ...699..513L} Li, Y.-R., Yuan, Y.-F.,
Wang, J.-M., Wang, J.-C., \& Zhang, S.\ 2009, \apj, 699, 513


\bibitem[Lister(2001)]{2001ApJ...562..208L} Lister, M.~L.\ 2001, \apj, 562,
208

\bibitem[Lister et al.(2009)]{2009AJ....137.3718L} Lister, M.~L., Aller,
H.~D., Aller, M.~F., et al.\ 2009, \aj, 137, 3718


\bibitem[Liu et al.(2002)]{2002ApJ...572L.173L} Liu, B.~F., Mineshige, S.,
\& Shibata, K.\ 2002, \apjl, 572, L173



\bibitem[Liu
\& Zhang(2011)]{2011ApJ...728L..44L} Liu, Y., \& Zhang, S.~N.\ 2011, \apjl, 728, L44


\bibitem[Lohfink et al.(2012)]{2012ApJ...758...67L} Lohfink, A.~M.,
Reynolds, C.~S., Miller, J.~M., et al.\ 2012, \apj, 758, 67

\bibitem[Malkan
\& Sargent(1982)]{1982ApJ...254...22M} Malkan, M.~A., \& Sargent,
W.~L.~W.\ 1982, \apj, 254, 22


\bibitem[Marconi et al.(2004)]{2004MNRAS.351..169M} Marconi, A., Risaliti,
G., Gilli, R., et al.\ 2004, \mnras, 351, 169

\bibitem[Marinucci et al.(2014)]{2014ApJ...787...83M} Marinucci, A., Matt, 
G., Miniutti, G., et al.\ 2014, \apj, 787, 83 

\bibitem[Martocchia et
al.(2002)]{2002A&A...387..215M} Martocchia, A., Matt, G., Karas, V., Belloni, T., \& Feroci, M.\ 2002, \aap, 387, 215


\bibitem[McClintock et al.(2011)]{2011CQGra..28k4009M} McClintock, J.~E.,
Narayan, R., Davis, S.~W., et al.\ 2011, Classical and Quantum Gravity, 28,
114009


\bibitem[McConnell et al.(2011)]{2011Natur.480..215M} McConnell, N.~J., Ma,
C.-P., Gebhardt, K., et al.\ 2011, \nat, 480, 215


\bibitem[Merloni(2004)]{2004MNRAS.353.1035M} Merloni, A.\ 2004, \mnras,
353, 1035

\bibitem[Meyer et
al.(2000)]{2000A&A...361..175M} Meyer, F., Liu, B.~F., \&
Meyer-Hofmeister, E.\ 2000, \aap, 361, 175




\bibitem[Miller et al.(2002)]{2002ApJ...577L..15M} Miller, J.~M., Fabian,
A.~C., in't Zand, J.~J.~M., et al.\ 2002, \apjl, 577, L15


\bibitem[Miniutti et al.(2004)]{2004MNRAS.351..466M} Miniutti, G., Fabian,
A.~C., \& Miller, J.~M.\ 2004, \mnras, 351, 466

\bibitem[Moderski
\& Sikora(1996)]{1996MNRAS.283..854M} Moderski, R., \& Sikora, M.\ 1996, \mnras, 283, 854

\bibitem[Molina et al.(2013)]{2013MNRAS.433.1687M} Molina, M., Bassani, L.,
Malizia, A., et al.\ 2013, \mnras, 433, 1687

\bibitem[Nandra et al.(1997)]{1997ApJ...488L..91N} Nandra, K., George,
I.~M., Mushotzky, R.~F., Turner, T.~J., \& Yaqoob, T.\ 1997, \apjl,
488, L91

\bibitem[Narayan
\& Yi(1994)]{1994ApJ...428L..13N} Narayan, R., \& Yi, I.\ 1994, \apjl, 428, L13


\bibitem[Novikov
\& Thorne(1973)]{1973blho.conf..343N} Novikov, I.~D., \& Thorne, K.~S.\ 1973, Black Holes (Les Astres Occlus), 343

\bibitem[Orosz et al.(2014)]{2014ApJ...794..154O} Orosz, J.~A., Steiner, 
J.~F., McClintock, J.~E., et al.\ 2014, \apj, 794, 154 

\bibitem[Page
\& Thorne(1974)]{1974ApJ...191..499P} Page, D.~N., \& Thorne, K.~S.\ 1974, \apj, 191, 499

\bibitem[Page et al.(2005)]{2005MNRAS.364..195P} Page, K.~L., Reeves,
J.~N., O'Brien, P.~T., \& Turner, M.~J.~L.\ 2005, \mnras, 364, 195


\bibitem[Pancoast et al.(2011)]{2011ApJ...730..139P} Pancoast, A., Brewer,
B.~J., \& Treu, T.\ 2011, \apj, 730, 139

\bibitem[Pancoast et al.(2012)]{2012ApJ...754...49P} Pancoast, A., Brewer,
B.~J., Treu, T., et al.\ 2012, \apj, 754, 49

\bibitem[Pancoast et al.(2014)]{2014MNRAS.445.3073P} Pancoast, A., Brewer, 
B.~J., Treu, T., et al.\ 2014, \mnras, 445, 3073 

\bibitem[Patrick et al.(2011)]{2011MNRAS.411.2353P} Patrick, A.~R., Reeves,
J.~N., Porquet, D., et al.\ 2011, \mnras, 411, 2353


\bibitem[Peterson et al.(2004)]{2004ApJ...613..682P} Peterson, B.~M.,
Ferrarese, L., Gilbert, K.~M., et al.\ 2004, \apj, 613, 682

\bibitem[Porquet et
al.(2004)]{2004A&A...422...85P} Porquet, D., Reeves, J.~N., O'Brien, P., \& Brinkmann, W.\ 2004, \aap, 422, 85

\bibitem[Qiao 
\& Liu(2015)]{2015MNRAS.448.1099Q} Qiao, E., \& Liu, B.~F.\ 2015, \mnras, 448, 1099 


\bibitem[Remillard
\& McClintock(2006)]{2006ARA&A..44...49R} Remillard, R.~A., \& McClintock, J.~E.\ 2006, \araa, 44, 49

\bibitem[Reynolds
\& Nowak(2003)]{2003PhR...377..389R} Reynolds, C.~S., \& Nowak, M.~A.\ 2003, \physrep, 377, 389


\bibitem[Runnoe et al.(2013)]{2013MNRAS.435.3251R} Runnoe, J.~C., Shang,
Z., \& Brotherton, M.~S.\ 2013, \mnras, 435, 3251


\bibitem[Sakimoto
\& Coroniti(1981)]{1981ApJ...247...19S} Sakimoto, P.~J., \& Coroniti, F.~V.\ 1981, \apj, 247, 19

\bibitem[Sambruna(1997)]{1997ApJ...487..536S} Sambruna, R.~M.\ 1997, \apj,
487, 536


\bibitem[Schlegel et al.(1998)]{1998ApJ...500..525S} Schlegel, D.~J.,
Finkbeiner, D.~P., \& Davis, M.\ 1998, \apj, 500, 525

\bibitem[Shafee et al.(2006)]{2006ApJ...636L.113S} Shafee, R., McClintock,
J.~E., Narayan, R., et al.\ 2006, \apjl, 636, L113


\bibitem[Shakura
\& Sunyaev(1973)]{1973A&A....24..337S} Shakura, N.~I., \& Sunyaev, R.~A.\ 1973, \aap, 24, 337


\bibitem[Shang et al.(2011)]{2011ApJS..196....2S} Shang, Z., Brotherton,
M.~S., Wills, B.~J., et al.\ 2011, \apjs, 196, 2

\bibitem[Shemmer et al.(2006)]{2006ApJ...646L..29S} Shemmer, O., Brandt,
W.~N., Netzer, H., Maiolino, R., \& Kaspi, S.\ 2006, \apjl, 646, L29


\bibitem[Shen et al.(2011)]{2011ApJS..194...45S} Shen, Y., Richards, G.~T.,
Strauss, M.~A., et al.\ 2011, \apjs, 194, 45


\bibitem[Shields(1978)]{1978Natur.272..706S} Shields, G.~A.\ 1978, \nat,
272, 706

\bibitem[Steiner et al.(2009a)]{2009ApJ...701L..83S} Steiner, J.~F.,
McClintock, J.~E., Remillard, R.~A., Narayan, R.,
\& Gou, L.\ 2009a, \apjl, 701, L83

\bibitem[Steiner et al.(2009b)]{2009PASP..121.1279S} Steiner, J.~F., 
Narayan, R., McClintock, J.~E., \& Ebisawa, K.\ 2009b, \pasp, 121, 1279 

\bibitem[Steiner et al.(2014)]{2014ApJ...793L..29S} Steiner, J.~F., 
McClintock, J.~E., Orosz, J.~A., et al.\ 2014, \apjl, 793, L29 

\bibitem[Stella
\& Rosner(1984)]{1984ApJ...277..312S} Stella, L., \& Rosner, R.\ 1984, \apj, 277, 312

\bibitem[Sun
\& Malkan(1989)]{1989ApJ...346...68S} Sun, W.-H., \& Malkan, M.~A.\ 1989, \apj, 346, 68


\bibitem[Sunyaev
\& Titarchuk(1985)]{1985A&A...143..374S} Sunyaev, R.~A., \& Titarchuk, L.~G.\ 1985, \aap, 143, 374


\bibitem[Svensson
\& Zdziarski(1994)]{1994ApJ...436..599S} Svensson, R., \& Zdziarski, A.~A.\ 1994, \apj, 436, 599

\bibitem[Taam
\& Lin(1984)]{1984ApJ...287..761T} Taam, R.~E., \& Lin, D.~N.~C.\ 1984, \apj, 287, 761


\bibitem[Tang et al.(2012)]{2012ApJS..201...38T} Tang, B., Shang, Z., Gu,
Q., Brotherton, M.~S., \& Runnoe, J.~C.\ 2012, \apjs, 201, 38


\bibitem[Vanden Berk et al.(2001)]{2001AJ....122..549V} Vanden Berk, D.~E.,
Richards, G.~T., Bauer, A., et al.\ 2001, \aj, 122, 549


\bibitem[Vasudevan
\& Fabian(2007)]{2007MNRAS.381.1235V} Vasudevan, R.~V., \& Fabian, A.~C.\ 2007, \mnras, 381, 1235


\bibitem[Vestergaard
\& Osmer(2009)]{2009ApJ...699..800V} Vestergaard, M., \& Osmer,
P.~S.\ 2009, \apj, 699, 800


\bibitem[Vestergaard
\& Peterson(2006)]{2006ApJ...641..689V} Vestergaard, M., \& Peterson, B.~M.\ 2006, \apj, 641, 689

\bibitem[Volonteri et al.(2005)]{2005ApJ...620...69V} Volonteri, M., Madau,
P., Quataert, E., \& Rees, M.~J.\ 2005, \apj, 620, 69


\bibitem[Wang et al.(2004)]{2004ApJ...607L.107W} Wang, J.-M., Watarai,
K.-Y., \& Mineshige, S.\ 2004, \apjl, 607, L107


\bibitem[Wills
\& Brotherton(1995)]{1995ApJ...448L..81W} Wills, B.~J., \& Brotherton, M.~S.\ 1995, \apjl, 448, L81


\bibitem[Woo
\& Urry(2002)]{2002ApJ...581L...5W} Woo, J.-H., \& Urry, C.~M.\ 2002, \apjl, 581, L5

\bibitem[Wu et al.(2013)]{2013ApJ...770...31W} Wu, Q., Cao, X., Ho, L.~C.,
\& Wang, D.-X.\ 2013, \apj, 770, 31

\bibitem[Wu et
al.(2004)]{2004A&A...424..793W} Wu, X.-B., Wang, R., Kong, M.~Z., Liu, F.~K., \& Han, J.~L.\ 2004, \aap, 424, 793

\bibitem[You et al.(2012)]{2012ApJ...761..109Y} You, B., Cao, X.,
\& Yuan, Y.-F.\ 2012, \apj, 761, 109

\bibitem[You et al.(2015)]{arXiv:1506.03959} You, B., Czerny, B., Sobolewska, M. et al.: 2015, arXiv:1506.03959

\bibitem[Yuan et al.(2010)]{2010ApJ...723..508Y} Yuan, W., Liu, B.~F.,
Zhou, H., \& Wang, T.~G.\ 2010, \apj, 723, 508

\bibitem[Zdziarski et
al.(1996)]{1996A&AS..120C.553Z} Zdziarski, A.~A., Gierlinski, M.,
Gondek, D., \& Magdziarz, P.\ 1996, \aaps, 120, 553


\bibitem[Zdziarski et al.(1999)]{1999MNRAS.303L..11Z} Zdziarski, A.~A.,
Lubi{\'n}ski, P., \& Smith, D.~A.\ 1999, \mnras, 303, L11

\bibitem[Zhang et
al.(1985)]{1985Ap&SS.113..181Z} Zhang, J.~L., Xiang, S.~P., \& Lu, J.~F.\ 1985, \apss, 113, 181


\bibitem[Zhang et al.(1997)]{1997ApJ...482L.155Z} Zhang, S.~N., Cui, W.,
\& Chen, W.\ 1997, \apjl, 482, L155


\bibitem[Zheng et al.(1997)]{1997ApJ...475..469Z} Zheng, W., Kriss, G.~A.,
Telfer, R.~C., Grimes, J.~P., \& Davidsen, A.~F.\ 1997, \apj, 475, 469

\bibitem[Zhou 
\& Zhao(2010)]{2010ApJ...720L.206Z} Zhou, X.-L., \& Zhao, Y.-H.\ 2010, \apjl, 720, L206 

\end{thebibliography}
\end{document}